 \documentclass[letter,useAMS,usenatbib]{mn2e}

 \usepackage[pdftex]{graphicx}

\def\msun{M$_\odot$}
\def\mstar{$M_\star$}

\def\ha{H$\alpha$}
\def\hb{H$\beta$}
\def\hc{H$\gamma$}
\def\hd{H$\delta$}
\def\oii{[OII]$\lambda$3727}
\def\nii{[NII]$\lambda$6584}
\def\neiii{[NeIII]$\lambda$3869}

\def\oiii{[OIII]$\lambda$4958,5007}

\def\mua{$\mu_\alpha$}
\def\muf{$\mu_{0.32}$}

\title[Long GRB hosts and the FMR]
{
The metallicity of the long GRB hosts 
and the Fundamental Metallicity Relation 
of low-mass galaxies
}

\author[F. Mannucci et al.]{
F. Mannucci$^1$\thanks{E-mail:filippo@arcetri.astro.it},
R. Salvaterra$^{2}$,
M. A. Campisi$^{2}$,\\
$^1$INAF - Osservatorio Astrofisico di Arcetri, 
   Largo E. Fermi 5, I-50125, Firenze, Italy\\
$^2$Dipartimento di Fisica e Matematica, Universit{\'a} dell'Insubria, via
Valleggio 7, 22100 Como, Italy
}

\begin{document}

\date{Submitted 2010 August}

\pagerange{\pageref{firstpage}--\pageref{lastpage}} \pubyear{2010}

\maketitle

\begin{abstract}
We investigate the metallicity properties of host galaxies of long Gamma-ray Bursts
(GRBs) in the light of the Fundamental Metallicity Relation (FMR), 
the tight dependence of metallicity on mass and SFR recently
discovered for SDSS galaxies with stellar masses above $10^{9.2}$\msun. 
As most of the GRB hosts have masses below this limit,
the FMR can only be used after an extension towards lower masses. 
At this aim, we study the FMR for galaxies with masses down to 
$\sim10^{8.3}$\msun, finding that the FMR does extend smoothly at lower masses,
albeit with a much larger scatter. 
We then compare the resulting FMR with the metallicity properties of 
18 host galaxies of long GRBs. While the GRB hosts show a systematic offset
with respect to the mass-metallicity relation, they are fully consistent with 
the FMR. This shows that the difference with the mass-metallicity 
relation is due to higher than average SFRs, and that 
GRBs  with optical afterglows do not preferentially select 
low-metallicity hosts 
among the star-forming galaxies. 
The apparent low metallicity is therefore a consequence of the
occurrence of long GRB in low mass, actively star-forming galaxies, known
to dominate the current cosmic SFR.
\end{abstract}

%
\section{Introduction}
\label{sec:intro}

Gamma-ray bursts (GRBs) are the most energetic explosions in
the Universe (see \citealt{Zhang04} for a review) and detected in the
$\gamma$-rays with a frequency of about one per day over the 
whole sky. The $\gamma-$ray emission is accompanied by a long-lasting
tail, called afterglow, usually detected over the whole electromagnetic
spectrum. Their extreme brightness easily over-shine the luminosity of
their host galaxy and makes them detectable up
to extreme high redshift as shown by the discovery of GRB~090423 at $z=8.2$
\citep{Salvaterra09,Tanvir09}. GRBs are usually 
divided in two, broad classes \citep{Kouveliotou93}:  short GRBs,
which are believed to result from the merger of two compact objects, and 
long GRBs, associated to the collapse of the core of a massive star, as a 
Wolf Rayet star \citep{Yoon06,Woosley06,yoon08}.
In this paper, we limit our analysis
to the class of long GRBs.

Recent studies on the final evolutionary stages of
massive stars \citep{Woosley06,Fryer99} have suggested that a Wolf-Rayet star can
produce a long GRB if its mass loss rate is small, which is
possible only if the metallicity of the star is lower than $\sim
0.1-0.3\,Z_{\odot}$. In this case, the specific angular momentum of the
progenitor allows the loss of the hydrogen envelope while
preserving the helium core. In this view, GRBs may occur preferentially
in galaxies with low-metallicity \citep{Fynbo03,Prochaska04,Fruchter06,Stanek06}, 
although we have to 
stress that low-metallicity progenitors do
not necessarily imply low-metallicity host galaxies. Indeed,
owing to the existence of metallicity gradients inside galaxies, GRBs could
form from low-metallicity progenitors also in hosts with relatively high 
metallicities \citep{Campisi09}.

Up to now, we have been able to detect the host galaxy of $\sim 70$ long GRBs
with known redshift.
In more than half of the cases, the observations allowed to
 determine the stellar mass and the star
formation rate of the 
galaxy\footnote{Data taken from http://www.grbhosts.org/, see \cite{Savaglio07}}.
The observational information gathered so far indicates that long GRBs 
with optical afterglows are typically
found to reside in low-mass, dwarf galaxies with average stellar masses 
$M_\star\sim 1-5\times 10^9$ \msun\ and high specific star formation rate 
(SSFR=SFR/$M_\star$). Information about the chemical content of this objects
are known only for a subsample of the hosts 
\citep{Savaglio09,Levesque10a,Levesque10c}. 
While most of the long GRBs are in low-metallicity
galaxies, a few cases for which the galaxy metallicity is found to be 
quite high do exist (e.g. GRB~020819, \citealt{Levesque10c,Kupcu-Yoldas10}, 
and GRB~050401, \citealt{Watson06}), so that 
the role of metallicity in driving the GRB phenomena remains unclear and 
it is still debated \citep{Fynbo03,Prochaska04,Stanek06,Fynbo06a,
Wolf07,Price07,Modjaz08,
Kocevski09,Savaglio09,Graham09a,Graham09c,Levesque10b,Levesque10c,Levesque10e,
Svensson10,Fan10b}.

Many recent studies  have attempted to find similarities and differences 
between the GRB host
population and the normal field galaxy one 
(see, for example, \citealt{Fynbo08}). In particular, these studies
compared the observed mass-metallicity relation (or luminosity-metallicity
relation) of the two populations obtaining contradictory results. From
the analysis of a whole sample of known GRB hosts, \cite{Savaglio09} 
concluded that there is no clear indication that GRB host galaxies belong
to a special population. Their properties are those expected for normal
star-forming galaxies, from the local to the most distant universe.
On the other hand, the study of sub-samples with well-determined chemical 
properties (e.g. \citealt{Levesque10b,Levesque10c,Han10a}) suggests that
most of the long GRB host galaxies fall below the $M-Z$ relation for
the normal galaxy population. \\

The aim of this work is to further test
the differences between GRB hosts and field galaxies
by taking advantage of the new Fundamental Metallicity Relation (FMR) 
recently introduced by \cite{Mannucci10}. 
The FMR is a tight  relation between stellar mass \mstar, SFR and 
gas-phase metallicity. Local SDSS galaxies define a surface in the 3D space
of these three quantities, with metallicity well determined by stellar mass 
and SFR. The residual metallicity scatter around this surface is very small,
about 0.05 dex, similar to the expected uncertainties. 
Also, the same FMR defined locally by SDSS galaxies is found 
to describe, without any evolution, the properties of high-redshift 
galaxies, up to z=2.5. The origin of the strong, monotonic evolution of 
the mass-metallicity relation over the same redshift range 
(e.g., \citealt{Tremonti04, Erb06b}) is due to the increase 
of target SFR with redshift,
resulting in sampling different parts of the same FMR at different redshifts.
At even higher redshifts, galaxies are found to evolve off the FMR 
\citep{Maiolino08,Mannucci09b}, and this effect is under test with a 
larger number of observations (Troncoso et al., in prep.).

The ranges of mass and SFR over which the FMR was measured are limited by 
the number of galaxies in the SDSS-DR7 sample used, which become rare 
at log(\mstar/\msun) below 9.2 and above 11.4, and 
at log(SFR/\msun\ yr$^{-1}$) below -1.4 and above +0.8.
For this reason, a comparison with the hosts of GRBs can only be done by 
extending this relation using lower mass galaxies, while a simple
extrapolation of the FMR of massive galaxies could produce spurious effects.


\begin{figure*}
\centerline{
   \includegraphics[angle=-90,width=0.95\textwidth]{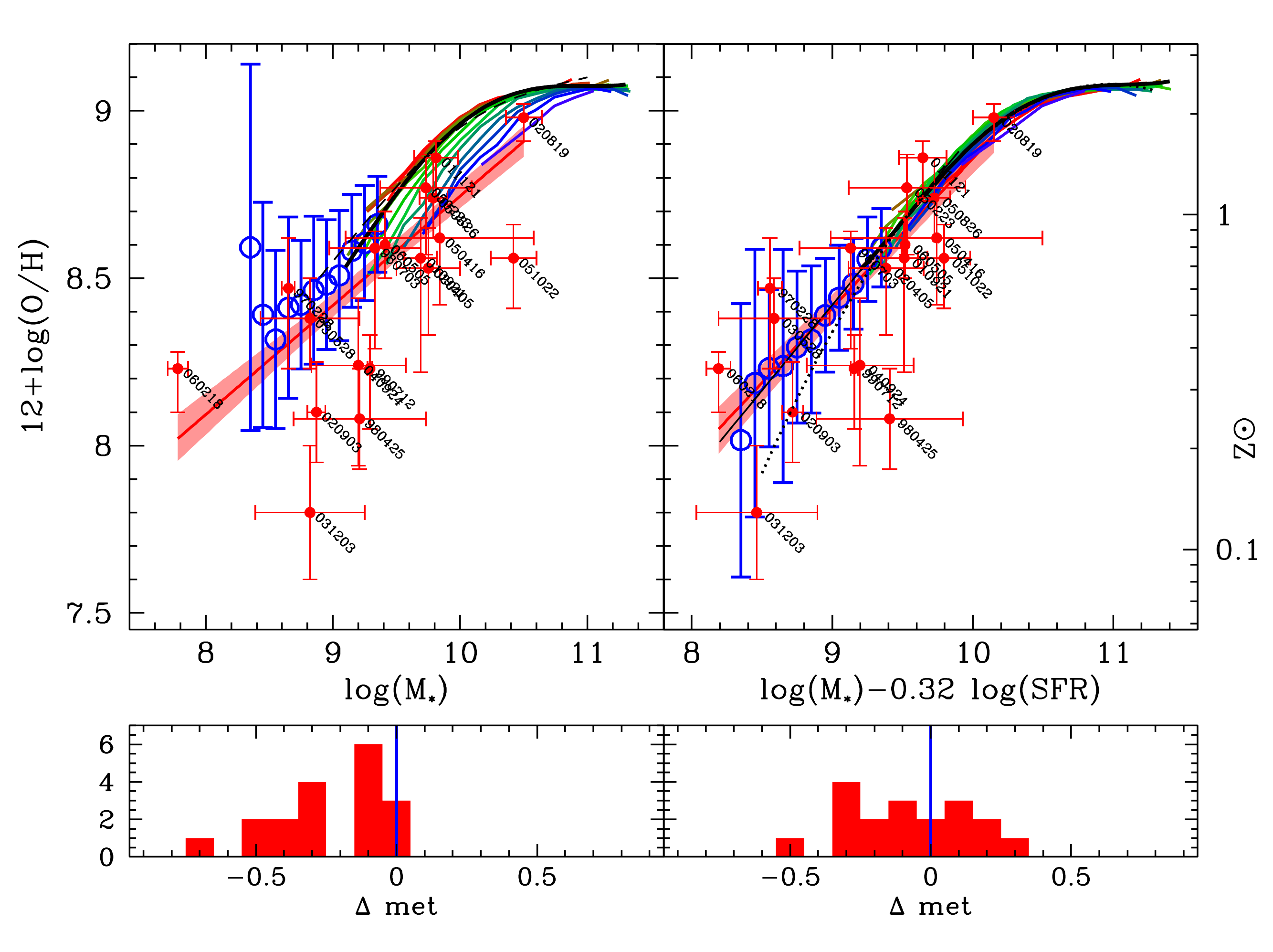} 
}
\caption{
{\em Left:}
Mass-Metallicity:, metallicity of low-mass SDSS galaxies 
(blue open dots with $1\sigma$ dispersions) 
as a function of stellar mass.
The coloured lines are local SDSS galaxies from Mannucci et al. (2010), color-coded
from red to blue according to increasing SFR. 
The black thick line shows the polynomial fit to the mass-metallicity relation in 
Mannucci et al. (2010). The black dashed line is the mass-metallicity 
relation in Jabran~Zahid et al. (2010), 
transformed to the same metallicity scale. 
The host galaxies of long GRBs are over-plotted (red solid dots, labelled with the GRB date). 
The red thick line is a linear fit to these GRB host data,
with $\pm1\sigma$ bands shown in light red. It is clear that GRB host follow a different relation
and show systematically lower metallicities. The {\em lower panel} shows the difference between 
the metallicity of the GRB hosts and the mass-metallicity relation of SDSS galaxies,
showing the systematic offset toward lower metallicities.
{\em Right:}
Fundamental metallicity relation:
metallicity as a function of $\mu_{0.32}$=log(\mstar)-0.32log(SFR) in solar units. 
The black solid line is the linear fit to the low-mass SDSS galaxies.
For comparison, the black dotted line is the extrapolation of the 2nd degree fit to 
the FMR of the SDSS galaxies as defined in Mannucci et al. (2010) and plotted for SFR=0. 
The linear fit to the GRB host data (red line with the $\pm1\sigma$ band)
shows that GRB hosts are fully compatible with the FMR defined by local SDSS galaxies.
This is also shown in the lower panel, where the metallicity difference of GRB hosts 
with the FMR is plotted.
}
\label{fig:grbmet}
\end{figure*}

\section{Extending the FMR towards smaller masses}
\label{sec:extend}

To derive the FMR, \cite{Mannucci10} have split $\sim$140000 SDSS-DR7 
galaxies into bins of 
mass and SFR having width of 0.15dex in both quantities. To have a good estimate 
of both median and dispersion of the metallicity 
for each value of mass and SFR, only bins containing more than 50 galaxies
have been used. This severely limits the range of mass and SFR over which 
the FMR has been measured, even if a significant number of galaxies outside 
these ranges are present in the original sample.
Among the $\sim$140000 SDSS-DR7 galaxies selected by \cite{Mannucci10} 
requiring 0.07$<$z$<$0.30 and signal-to-noise ratio SNR(\ha)$>$25, 
about 2000 ($1.4\%$)
have masses below $10^{9.2}$\msun. Here we intend to use these 
galaxies to extend the measured FMR.\\

\cite{Mannucci10} have introduced the new quantity \mua\ defined as a linear 
combination of stellar mass and SFR:
\begin{equation}
\mu_\alpha={\rm log(M_*)}-\alpha\ \rm{log(SFR)}
\end{equation}

\noindent
and have demonstrated that, for $\alpha$=0.32,
all galaxies at z$<$2.5 show the same dependence of metallicity 
on \muf\ and the same range of values of \muf. In other words, the introduction 
of this quantity roughly defines a projection of the FMR that minimizes 
the scatter, i.e, corresponds to observing the FMR "edge on".
From a physical point of view, metallicity is found to increase with 
mass and decrease 
with SFR, therefore a combination of these two quantities, with a negative factor 
for SFR, is expected, and found, to show a better correlation with metallicity.
It is worth noticing that the dependence of metallicity only on \muf\ is not exact, 
as no part of the FMR is exactly a plane (see Fig. 2 of \citealt{Mannucci10}), 
nevertheless this is a convenient approximation.

To avoid binning the limited number of galaxies with low mass
into a large number of classes of mass and SFR, 
we extend the FMR directly by considering the combination \muf. 
We consider the $\sim$1400 galaxies in \cite{Mannucci10} sample with 8.3$<$\muf$<$9.4. 
This is a small sample, side by side to the large sample of $\sim$140.000 galaxies 
with larger values of \muf, and problems with contamination are possible.
Indeed, while $\sim$1300 galaxies with low \muf\ show low metallicities, 86 of them, 
corresponding to 0.0007 of the full sample, 
have large values of metallicities, above 12+log(O/H)=8.9, with
the same distribution of metallicity of 
the large population of massive, quiescent, metal-rich galaxies. 
Given the intrinsic uncertainties on mass and SFR, these 86 galaxies
are likely to be metal-rich galaxies whose \muf\ 
is incorrectly measured and scattered towards low values. 
We remove these galaxies from the sample, and
divide the remaining $\sim$1300 galaxies in bins of \muf, 
and for each bin we compute median and standard deviation of metallicity.
For comparison, we also compute the mass-metallicity relation considering the $\sim$1700
galaxies with masses between $10^{8.3}$ and $10^{9.4}$\msun.

The results are shown in figure~\ref{fig:grbmet}, where low-mass galaxies are 
compared to galaxies of larger \mstar. 
The left panel shows the mass-metallicity relation. 
At masses above $\sim10^{10}$\msun\ this is fully consistent 
with \cite{Tremonti04}, while it 
shows lower values of metallicity and a steeper slope at lower masses. 
This is probably due to the different selections: our requirement of a high signal-to-noise ratio
(SNR$>$25, see \citealt{Mannucci10}) in the \ha\ line preferentially select galaxies with high
SFR and, as a consequence, lower metallicity, especially at low stellar masses.

On the right of figure~\ref{fig:grbmet}, the FMR is shown.
Low mass galaxies appear to extend smoothly the FMR, 
with a linear relation between metallicity and \muf. 
The resulting FMR can be described by:

\begin{equation}
\begin{array}{rl}
12+log(O/H)&=8.90+0.37m-0.14s-0.19m^2                    \\
           &+0.12ms-0.054s^2~~~~~~~~~~\rm{for}~\mu_{0.32}\ge9.5 \\
           &=8.93+0.51(\mu_{0.32}-10)~~\rm{for}~\mu_{0.32}<9.5 \\
\end{array}
\end{equation}
where $m$=log(\mstar)-10 and $s$=log(SFR) in solar units.

It is evident that the intrinsic scatter around the FMR increases towards
lower values of \muf. The residual scatter is larger than the expected 
errors on metallicity, mass and SFR, even if
the uncertainties on stellar masses from SED fitting could increase 
towards low masses.  This increasing scatter
towards dwarf galaxies is a well known effect probably related to a large
spread of histories and current levels of star formation 
\citep{Hunter99,Hunt05a,Zhao10}. 

\section{The metallicity of the hosts of GRB}
\label{sec:hosts}

The formation of long GRBs is thought to be related to the collapse of a 
very massive, low metallicity star \citep{Woosley06,Fryer99}.
Thus, it has been argued that the occurrence of a long GRB may be linked
to a overall low metal content of its host galaxy, 
making GRB hosts a biased galaxy sample with respect to the normal field 
population. 
In order to check whether this bias exists, we consider the properties of the 
GRB hosts in the light of the observed FMR for normal field galaxies.
To this extent, we collect all the GRB host 
galaxies at z$<$1 with available observations
to measure, at the same time, 
stellar mass, SFR, and gas phase metallicity.

Line fluxes of long GRB hosts have been published by several authors 
\citep{Savaglio09,Han10a,Levesque10d}. We have used these compilations to measure 
gas-phase metallicities, using the method described in \cite{Maiolino08} 
and \cite{Cresci10}, and expressing it in the same scale 
as in \cite{Nagao06}, \cite{Kewley08}, 
and \cite{Mannucci10}, where solar metallicity is 12+log(O/H)=8.69. 
Many metallicity indicators have been proposed that are based on line ratios
(e.g., \citealt{Pettini04,Nagao06,Kewley08}), but none of them is 
without problem.
For example, R23 has two branches, with two different metallicities associated 
to each value of R23. Both \ha/\nii\ and \oii/\neiii\ have monotonic variations 
with metallicity and no dependence on extinction, but they include fainter lines,
especially for very high or very-low metallicities. 
\oiii/\oii\ and \oiii/\nii\ are sensitive to extinction, which is usually 
poorly known.
Following \cite{Nagao06}, we measure metallicities by simultaneously considering all 
the flux ratios among the relevant emission lines, fitting these values
with two free parameters, metallicity and extinction. Usually this method  
can obtain a reliable value of metallicity, avoiding or mitigating the intrinsic 
problems of each individual line ratio. In contrast, extinction is usually
very poorly constrained, because most of the line ratios used involve line with 
similar wavelengths. 
For this reason, when flux ratios
between different hydrogen Balmer lines are available and give consistent results, 
we measure extinction from these Balmer decrement (assuming intrinsic line ratios of
\ha/\hb=2.87, \hc/\hb=0.466, \hd/\hb=0.256, \citealt{Osterbrock89}), 
considerably reducing the uncertainties on $A_V$. 
The SFR is then obtained, 
as in \cite{Mannucci10}, from \ha\ corrected for extinction, 
using the calibration in \cite{Kennicutt98}. 
Uncertainties on the SFR are computed taking into account the errors on 
both line fluxes and dust extinction.
Finally 
stellar masses are taken from  \cite{Savaglio09}.
Table~1 lists the resulting properties of the host galaxies
in terms of stellar mass, SFR, gas-phase metallicity, and intrinsic dust extinction.

These data are plotted in fig.~\ref{fig:grbmet} and compared with both
the mass-metallicity relation (left panel) and the FMR (right panel) of local SDSS galaxies.
We computed a linear fit to the GRB host data taking into account the errors on 
metallicity, mass and SFR.
The comparison with the mass-metallicity relation shows that,
as already obtained by \cite{Levesque10d} and \cite{Han10a}, GRB host galaxies have lower 
metallicity than galaxies of the same mass
both in the local universe (SDSS galaxies) and at intermediate redshift 
\citep{Savaglio05, Jabran-Zahid10}. In contrast, we also find that GRB hosts
do follow the FMR and its extension towards low masses, without any significant discrepancy.
In other words, when the dependence on SFR is properly taken into account,
the metallicity properties of long GRB hosts do not differ substantially from
those of the typical field population. 
As explained in the discussion, this means that
the low metallicities are associated to both low masses and high SFR, i.e,
to high SSFR.

We stress that such a good agreement is only obtained when the original
FMR is extended using low-mass galaxies. The use of an extrapolation of the original 
2nd-order fit would produce a spurious difference in metallicity, 
with GRB hosts more metal rich than field galaxies.

\begin{figure}
\centerline{
   \includegraphics[width=0.48\textwidth]{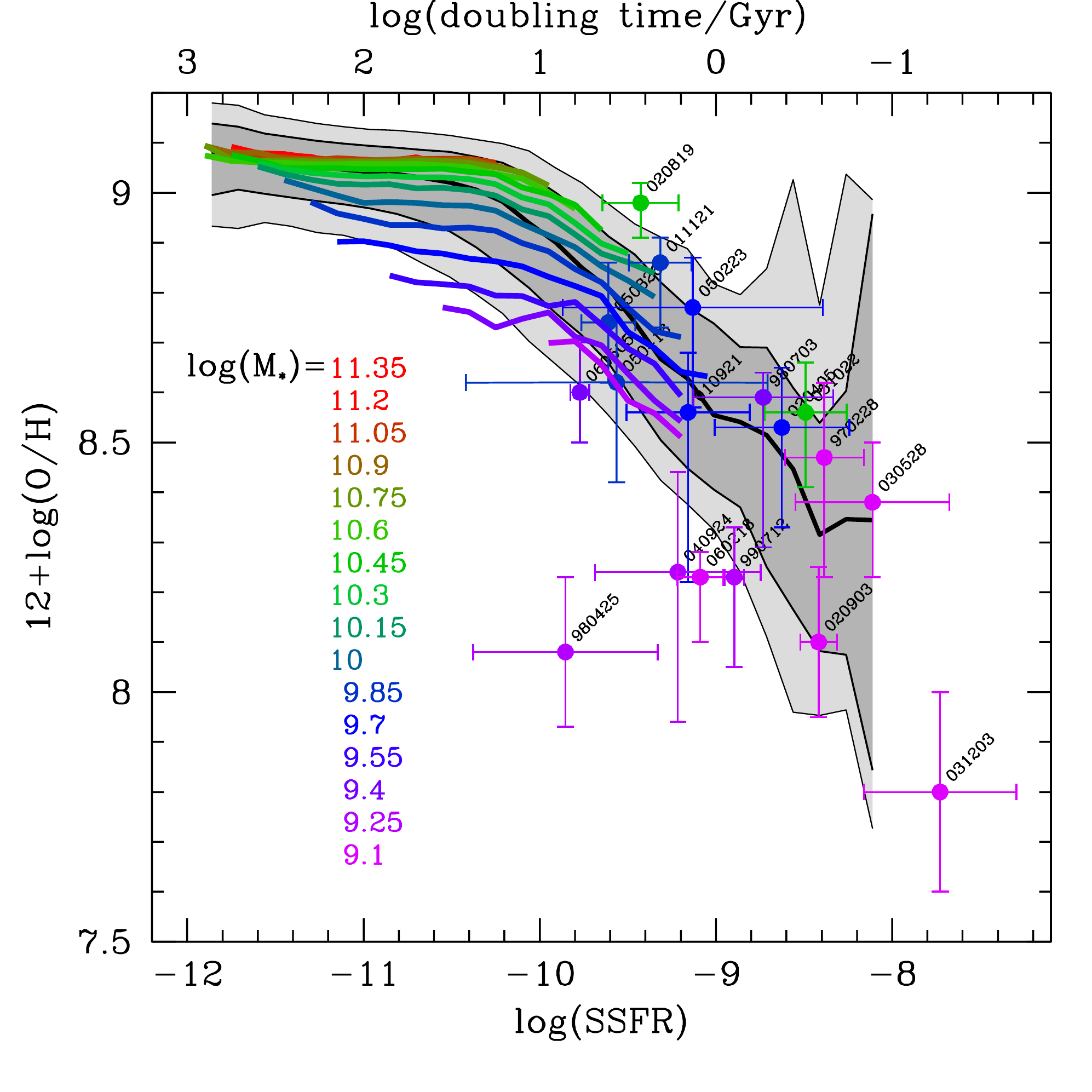} 
}
\caption{
Metallicity of GRB hosts as a function of the SSFR. 
The grey-shaded area shows the relation for all SDSS galaxies, irrespective of mass, 
with the areas containing 68\% and 90\% of the galaxies. 
The colored lines are the median metallicity
of SDSS galaxies with the listed values of stellar mass. Dots are GRB hosts, color coded 
with stellar mass. A broad agreement between the two distributions is found.
}
\label{fig:ssfr}
\end{figure}

In figure~\ref{fig:ssfr} we plot the relation between SSFR 
and metallicity for the 18 GRB host galaxies of our sample
compared to the local SDSS galaxies.
Here the color code shows different values of stellar mass. 
The solid lines show the relation between mass and SSFR for
field galaxies, and the shaded area accounts for the intrinsic scatter 
of the observed relation for SDSS galaxies. GRB hosts  
populate the plot similarly to normal field galaxy population, with more
(less) massive hosts lying in the upper (lower) bound of the observed 
relation. As already discussed, host metallicities are in line with 
those expected
for star forming, field objects, apart GRB~980425.
Notably, all the GRB hosts are
found to present relatively high SSFR, with log(SSFR)$\ge -10$. Their
growth time, i.e. the time required by the galaxy to form its observed 
stellar mass at the present level of SFR, i.e. 1/SSFR, 
is shorter than the Hubble time at the redshift of the GRB, for
all objects in our sample. This indicates that GRB host are forming quite 
efficiently their stars similarly to local starbursts.

\begin{table}
\label{tab:hosts}
\caption{Properties of the GRB hosts}
\renewcommand{\tabcolsep}{3pt}
\begin{tabular}{llrccc}
\hline
\hline
  GRB    & z      &   log(\mstar)  &   SFR          &  12+log(O/H)           &  $A_V$  \\
         &        &    (\msun)     &  (\msun/yr)    &                        &         \\
\hline
 970228  & 0.695  & 8.65$\pm$0.05 & 1.95 $\pm$0.22 & $8.47^{+0.15}_{-0.24}$ & 0.0$^{+0.7}_{-0.0}$\\
 980425  & 0.0085 & 9.21$\pm$0.52 & 0.24 $\pm$0.05 & $8.08^{+0.15}_{-0.15}$ & 1.9$^{+0.1}_{-0.1}$\\
 980703  & 0.966  & 9.33$\pm$0.36 & 4.20 $\pm$0.17 & $8.59^{+0.05}_{-0.30}$ & 0.0$^{+0.7}_{-0.0}$\\
 990712  & 0.434  & 9.29$\pm$0.02 & 2.62 $\pm$0.05 & $8.23^{+0.10}_{-0.18}$ & 0.5$^{+0.1}_{-0.1}$\\
 010921  & 0.451  & 9.69$\pm$0.13 & 3.60 $\pm$0.32 & $8.56^{+0.12}_{-0.34}$ & 1.6$^{+1.0}_{-1.0}$\\
 011121  & 0.362  & 9.81$\pm$0.17 & 3.30 $\pm$0.05 & $8.86^{+0.05}_{-0.13}$ & 0.9$^{+0.1}_{-0.1}$\\
 020405  & 0.691  & 9.75$\pm$0.25 & 14.1 $\pm$0.29 & $8.53^{+0.12}_{-0.20}$ & 1.9$^{+0.6}_{-0.6}$\\
 020819B & 0.411  &10.50$\pm$0.14 & 12.5 $\pm$0.17 & $8.98^{+0.07}_{-0.07}$ & 1.8$^{+0.5}_{-0.5}$\\
 020903  & 0.251  & 8.87$\pm$0.07 & 3.00 $\pm$0.08 & $8.05^{+0.16}_{-0.15}$ & 0.8$^{+0.2}_{-0.2}$\\
 030528  & 0.782  & 8.82$\pm$0.39 & 5.40 $\pm$0.19 & $8.38^{+0.12}_{-0.15}$ & 0.0$^{+0.8}_{-0.0}$\\
 031203  & 0.1055 & 8.82$\pm$0.43 & 13.1 $\pm$0.05 & $7.80^{+0.20}_{-0.20}$ & 0.0$^{+0.2}_{-0.0}$\\
 040924  & 0.858  & 9.20$\pm$0.37 & 1.02 $\pm$0.58 & $8.23^{+0.20}_{-0.30}$ & 0.0$^{+1.2}_{-0.0}$\\
 050223  & 0.584  & 9.73$\pm$0.36 & 4.20 $\pm$0.64 & $8.77^{+0.10}_{-0.20}$ & 1.5$^{+1.4}_{-1.3}$\\
 050416  & 0.6528 & 9.84$\pm$0.74 & 2.00 $\pm$0.43 & $8.62^{+0.12}_{-0.20}$ & 0.7$^{+1.1}_{-0.7}$\\
 050826  & 0.296  & 9.79$\pm$0.11 & 1.60 $\pm$0.10 & $8.74^{+0.12}_{-0.12}$ & 0.1$^{+0.2}_{-0.1}$\\
 051022  & 0.8070 &10.42$\pm$0.18 & 89.6 $\pm$0.15 & $8.56^{+0.10}_{-0.15}$ & 1.0$^{+0.3}_{-0.3}$\\
 060218  & 0.0334 & 7.78$\pm$0.08 & 0.052$\pm$0.10 & $8.23^{+0.05}_{-0.13}$ & 0.5$^{+0.3}_{-0.3}$\\
 060505  & 0.0889 & 9.41$\pm$0.01 & 0.46 $\pm$0.05 & $8.60^{+0.10}_{-0.10}$ & 0.8$^{+0.1}_{-0.1}$\\
\hline
\end{tabular}
\end{table}

\section{Discussion}

We have compared the metallicity properties of a sample of 18 GRB host galaxies with
those of the local field population. In particular, we have found that
GRB hosts do follow the FMR recently found by \cite{Mannucci10}. This
fact implies that GRB hosts do not differ substantially from the typical
galaxy population. 
The typical low, sub-solar  metallicity found in many recent studies
(e.g., \citealt{Savaglio09,Levesque10d} and references therein) 
does not necessary mean that GRBs occur in special, low-metallicity galaxies, 
as the exception of GRB~020819 clearly shows, 
and that a direct link between low metallicity and GRB production exists. 
Indeed, this observation can be explained as a consequence of the well-known
link between the GRB event and the death of a very massive stars, 
which produces a relation between long GRBs and star formation
\citep{Totani97,Mao98,Wijers98,Porciani01}.
In the local universe, about 70\% of all star formation activity occurs in 
 galaxies with masses between $10^{9.5}$ and $10^{10.2}$\msun
\citep{Mobasher09}, where most of the GRB of our sample are also found. 
Low stellar mass means low metallicities at all redshifts 
(e.g., \citealt{Tremonti04,Savaglio05,Jabran-Zahid10,Erb06a,Maiolino08,Mannucci09b}),
therefore low metallicities are expected for GRB hosts.
Also, the FMR shows that galaxies with higher SFR also have lower metallicities
than more quiescent galaxies of the same mass. As a consequence, a star-formation
selected galaxy sample, such as the present GRB host sample, is expected to
fall below the mass-metallicity relation, but follow the FMR. 
This is what is observed in figure~\ref{fig:grbmet}.

Some warnings apply, which are related to the nature of the
present GRB host galaxy sample.
Our sample consists mostly of long GRBs whose position has been provided 
by the detection of their optical afterglow. It is known that a population of 
GRBs with a bright X-ray afterglow and lacking of optical counterpart 
does exist, the so-called dark GRBs, and most of them
reside in dusty environments (e.g., \citealt{Perley09b,Kupcu-Yoldas10}).
It is still not known if this dust is spread across the host galaxy, 
which thus is likely to be metal-rich \citep{Fynbo09}, or is directly associated to the 
GRB itself \citep{Perley09b}, maybe without a clear dependece on metallicity.
In all cases, it is
difficult to access with the available data whether the inclusion of dark GRBs
could change our main conclusions. Indeed, it is possible that 
dark GRB hosts would populate the region of the FMR at high value of $\mu_\alpha$.
This kind of studies will require the collection of an unbiased
GRB-selected galaxy sample (see e.g. \citealt{Malesani09}). 

Such a complete or well controlled sample is also needed to address the 
role of the several selection effects that could exist even within the 
class of GRBs with an optical bright afterglow. 
For example, it is very probable that galaxies with high SFR are over-represented 
in this sample because they are easier to detect.  
For this reason, for example, 
the present sample of GRB hosts does not allow us to study if 
a direct correspondence exist between the fraction of GRB and the fraction 
of SFR as a function of stellar mass of the hosts. 
Despite this problem, it is likely that the present sample represents the
metallicity properties of the population of host of GRBs 
with optical bright afterglows because none of the main
selection effects  within this class of GRBs 
is directly related to metallicity.

\section{Conclusions}

Our main findings can be summarized in the following two points: \\ 
(I)
 the average low metallicity observed
in  long GRB host galaxies is an outcome of the observed relation between 
stellar mass, SFR and metallicity: since the  long GRB hosts are generally 
low mass and high star forming galaxies, i.e. objects characterized by a 
low value of $\mu_{\alpha}$, their metallicity is expected to be sub-solar;
the metallicity observed in GRB hosts is exactly what is expected 
on the basis on their mass and SFR, with no apparent bias towards 
lower metallicities.\\
(II)
 long GRBs not necessary explode in galaxies with a low metallicity (and indeed 
GRB~020819 is one of such cases). The condition for a galaxy to host a GRB
seems indeed related to its ability to form stars in a efficient way. 
Our  conclusions,  based on a sample of GRB host at z$<$1, are similar to what 
has been found by \cite{Fynbo08} at high redshift.

Also, since GRB hosts follow the FMR, the relation can be used to
predict the properties of those hosts for which one of the parameters (\mstar,
SFR or metallicity) is not known. 

Finally, our results also suggest that larger samples of GRB hosts
can be used to study the FMR of normal starburst galaxies.
This is in particular exciting since GRBs may allow to
extend current studies of the FMR both toward low values of $\mu_\alpha$ and 
toward higher redshift, in principle up to 
extremely high redshifts (at least up to $z\sim 8$ as shown by GRB 090423).
Thank to they brightness, GRBs can be used as a background light to study the
metal content of its parent galaxy even at very high-$z$ as demonstrated by
the case of GRB~050904 at $z=6.3$ for which an estimate of the metallicity
has been derived by \cite{Totani06} and \cite{Kawai06}. The study of the metal enrichment history
at these early cosmic epochs is of uttermost importance to better understand
the first stages of galaxy formation in the Universe and to constrain the
properties of those galaxies that have re-ionized the intergalactic medium 
(see \citealt{Salvaterra10}).\\

\noindent
{\bf Acknowledgements}

We thank the Sandra Savaglio and the GHosts team for having collected the data on GRB hosts
in the GHostS database (www.grbhosts.org), which is 
partly funded by Spitzer/NASA grant RSA Agreement No. 1287913.
FM acknowledges partial financial support of the Italian Space Agency 
through contracts ASI-INAF I/016/07/0 and I/009/10/0, and of PRIN-INAF 2008.
We also thank the referee for useful discussions.

\bibliographystyle{/Users/filippo/arcetri/Papers/aa-package/bibtex/aa}
\bibliography{/Users/filippo/arcetri/bibdesk/Bibliography}

\begin{thebibliography}{61}
\expandafter\ifx\csname natexlab\endcsname\relax\def\natexlab#1{#1}\fi

\bibitem[{{Campisi} {et~al.}(2009){Campisi}, {De Lucia}, {Li}, {Mao}, \&
  {Kang}}]{Campisi09}
{Campisi}, M.~A., {De Lucia}, G., {Li}, L., {Mao}, S., \& {Kang}, X. 2009,
  \mnras, 400, 1613

\bibitem[{{Cresci} {et~al.}(2010){Cresci}, {Mannucci}, {Maiolino}, {Marconi},
  {Gnerucci}, \& {Magrini}}]{Cresci10}
{Cresci}, G., {Mannucci}, F., {Maiolino}, R., {et~al.} 2010, \nat, 467, 811

\bibitem[{{Erb} {et~al.}(2006{\natexlab{a}}){Erb}, {Shapley}, {Pettini},
  {Steidel}, {Reddy}, \& {Adelberger}}]{Erb06a}
{Erb}, D.~K., {Shapley}, A.~E., {Pettini}, M., {et~al.} 2006{\natexlab{a}},
  \apj, 644, 813

\bibitem[{{Erb} {et~al.}(2006{\natexlab{b}}){Erb}, {Steidel}, {Shapley},
  {Pettini}, {Reddy}, \& {Adelberger}}]{Erb06b}
{Erb}, D.~K., {Steidel}, C.~C., {Shapley}, A.~E., {et~al.} 2006{\natexlab{b}},
  \apj, 647, 128

\bibitem[{{Fan} {et~al.}(2010){Fan}, {Yin}, \& {Matteucci}}]{Fan10b}
{Fan}, X.~L., {Yin}, J., \& {Matteucci}, F. 2010, \aap, 521, A73+

\bibitem[{{Fruchter} {et~al.}(2006){Fruchter}, {Levan}, {Strolger},
  {Vreeswijk}, {Thorsett}, {Bersier}, {Burud}, {Castro Cer{\'o}n},
  {Castro-Tirado}, {Conselice}, {Dahlen}, {Ferguson}, {Fynbo}, {Garnavich},
  {Gibbons}, {Gorosabel}, {Gull}, {Hjorth}, {Holland}, {Kouveliotou}, {Levay},
  {Livio}, {Metzger}, {Nugent}, {Petro}, {Pian}, {Rhoads}, {Riess}, {Sahu},
  {Smette}, {Tanvir}, {Wijers}, \& {Woosley}}]{Fruchter06}
{Fruchter}, A.~S., {Levan}, A.~J., {Strolger}, L., {et~al.} 2006, \nat, 441,
  463

\bibitem[{{Fryer} {et~al.}(1999){Fryer}, {Woosley}, \& {Hartmann}}]{Fryer99}
{Fryer}, C.~L., {Woosley}, S.~E., \& {Hartmann}, D.~H. 1999, \apj, 526, 152

\bibitem[{{Fynbo} {et~al.}(2003){Fynbo}, {Jakobsson}, {M{\"o}ller}, {Hjorth},
  {Thomsen}, {Andersen}, {Fruchter}, {Gorosabel}, {Holland}, {Ledoux},
  {Pedersen}, {Rhoads}, {Weidinger}, \& {Wijers}}]{Fynbo03}
{Fynbo}, J.~P.~U., {Jakobsson}, P., {M{\"o}ller}, P., {et~al.} 2003, \aap, 406,
  L63

\bibitem[{{Fynbo} {et~al.}(2009){Fynbo}, {Jakobsson}, {Prochaska}, {Malesani},
  {Ledoux}, {de Ugarte Postigo}, {Nardini}, {Vreeswijk}, {Wiersema}, {Hjorth},
  {Sollerman}, {Chen}, {Th{\"o}ne}, {Bj{\"o}rnsson}, {Bloom}, {Castro-Tirado},
  {Christensen}, {De Cia}, {Fruchter}, {Gorosabel}, {Graham}, {Jaunsen},
  {Jensen}, {Kann}, {Kouveliotou}, {Levan}, {Maund}, {Masetti},
  {Milvang-Jensen}, {Palazzi}, {Perley}, {Pian}, {Rol}, {Schady}, {Starling},
  {Tanvir}, {Watson}, {Xu}, {Augusteijn}, {Grundahl}, {Telting}, \&
  {Quirion}}]{Fynbo09}
{Fynbo}, J.~P.~U., {Jakobsson}, P., {Prochaska}, J.~X., {et~al.} 2009, \apjs,
  185, 526

\bibitem[{{Fynbo} {et~al.}(2008){Fynbo}, {Prochaska}, {Sommer-Larsen},
  {Dessauges-Zavadsky}, \& {M{\o}ller}}]{Fynbo08}
{Fynbo}, J.~P.~U., {Prochaska}, J.~X., {Sommer-Larsen}, J.,
  {Dessauges-Zavadsky}, M., \& {M{\o}ller}, P. 2008, \apj, 683, 321

\bibitem[{{Fynbo} {et~al.}(2006){Fynbo}, {Starling}, {Ledoux}, {Wiersema},
  {Th{\"o}ne}, {Sollerman}, {Jakobsson}, {Hjorth}, {Watson}, {Vreeswijk},
  {M{\o}ller}, {Rol}, {Gorosabel}, {N{\"a}r{\"a}nen}, {Wijers},
  {Bj{\"o}rnsson}, {Castro Cer{\'o}n}, {Curran}, {Hartmann}, {Holland},
  {Jensen}, {Levan}, {Limousin}, {Kouveliotou}, {Nelemans}, {Pedersen},
  {Priddey}, \& {Tanvir}}]{Fynbo06a}
{Fynbo}, J.~P.~U., {Starling}, R.~L.~C., {Ledoux}, C., {et~al.} 2006, \aap,
  451, L47

\bibitem[{{Graham} {et~al.}(2009{\natexlab{a}}){Graham}, {Fruchter}, {Kewley},
  {Levesque}, {Levan}, {Tanvir}, {Reichart}, \& {Nysewander}}]{Graham09c}
{Graham}, J.~F., {Fruchter}, A.~S., {Kewley}, L.~J., {et~al.}
  2009{\natexlab{a}}, in American Institute of Physics Conference Series, Vol.
  1133, American Institute of Physics Conference Series, ed. {C.~Meegan,
  C.~Kouveliotou, \& N.~Gehrels}, 269--272

\bibitem[{{Graham} {et~al.}(2009{\natexlab{b}}){Graham}, {Fruchter}, {Levan},
  {Melandri}, {Kewley}, {Levesque}, {Nysewander}, {Tanvir}, {Dahlen},
  {Bersier}, {Wiersema}, {Bonfield}, \& {Martinez-Sansigre}}]{Graham09a}
{Graham}, J.~F., {Fruchter}, A.~S., {Levan}, A.~J., {et~al.}
  2009{\natexlab{b}}, \apj, 698, 1620

\bibitem[{{Han} {et~al.}(2010){Han}, {Hammer}, {Liang}, {Flores}, {Rodrigues},
  {Hou}, \& {Wei}}]{Han10a}
{Han}, X.~H., {Hammer}, F., {Liang}, Y.~C., {et~al.} 2010, \aap, 514, A24+

\bibitem[{{Hunt} {et~al.}(2005){Hunt}, {Bianchi}, \& {Maiolino}}]{Hunt05a}
{Hunt}, L., {Bianchi}, S., \& {Maiolino}, R. 2005, \aap, 434, 849

\bibitem[{{Hunter} \& {Hoffman}(1999)}]{Hunter99}
{Hunter}, D.~A. \& {Hoffman}, L. 1999, \aj, 117, 2789

\bibitem[{{Jabran Zahid} {et~al.}(2010){Jabran Zahid}, {Kewley}, \&
  {Bresolin}}]{Jabran-Zahid10}
{Jabran Zahid}, H., {Kewley}, L.~J., \& {Bresolin}, F. 2010, ArXiv e-prints
  1006.4877

\bibitem[{{Kawai} {et~al.}(2006){Kawai}, {Kosugi}, {Aoki}, {Yamada}, {Totani},
  {Ohta}, {Iye}, {Hattori}, {Aoki}, {Furusawa}, {Hurley}, {Kawabata},
  {Kobayashi}, {Komiyama}, {Mizumoto}, {Nomoto}, {Noumaru}, {Ogasawara},
  {Sato}, {Sekiguchi}, {Shirasaki}, {Suzuki}, {Takata}, {Tamagawa}, {Terada},
  {Watanabe}, {Yatsu}, \& {Yoshida}}]{Kawai06}
{Kawai}, N., {Kosugi}, G., {Aoki}, K., {et~al.} 2006, \nat, 440, 184

\bibitem[{{Kennicutt}(1998)}]{Kennicutt98}
{Kennicutt}, Jr., R.~C. 1998, \araa, 36, 189

\bibitem[{{Kewley} \& {Ellison}(2008)}]{Kewley08}
{Kewley}, L.~J. \& {Ellison}, S.~L. 2008, \apj, 681, 1183

\bibitem[{{Kocevski} {et~al.}(2009){Kocevski}, {West}, \&
  {Modjaz}}]{Kocevski09}
{Kocevski}, D., {West}, A.~A., \& {Modjaz}, M. 2009, \apj, 702, 377

\bibitem[{{Kouveliotou} {et~al.}(1993){Kouveliotou}, {Meegan}, {Fishman},
  {Bhat}, {Briggs}, {Koshut}, {Paciesas}, \& {Pendleton}}]{Kouveliotou93}
{Kouveliotou}, C., {Meegan}, C.~A., {Fishman}, G.~J., {et~al.} 1993, \apjl,
  413, L101

\bibitem[{{K{\"u}pc{\"u} Yolda{\c s}} {et~al.}(2010){K{\"u}pc{\"u} Yolda{\c
  s}}, {Greiner}, {Klose}, {Kr{\"u}hler}, \& {Savaglio}}]{Kupcu-Yoldas10}
{K{\"u}pc{\"u} Yolda{\c s}}, A., {Greiner}, J., {Klose}, S., {Kr{\"u}hler}, T.,
  \& {Savaglio}, S. 2010, \aap, 515, L2+

\bibitem[{{Levesque} {et~al.}(2010{\natexlab{a}}){Levesque}, {Berger},
  {Kewley}, \& {Bagley}}]{Levesque10b}
{Levesque}, E.~M., {Berger}, E., {Kewley}, L.~J., \& {Bagley}, M.~M.
  2010{\natexlab{a}}, \aj, 139, 694

\bibitem[{{Levesque} {et~al.}(2010{\natexlab{b}}){Levesque}, {Kewley},
  {Berger}, \& {Jabran Zahid}}]{Levesque10d}
{Levesque}, E.~M., {Kewley}, L.~J., {Berger}, E., \& {Jabran Zahid}, H.
  2010{\natexlab{b}}, \aj, 140, 1557

\bibitem[{{Levesque} {et~al.}(2010{\natexlab{c}}){Levesque}, {Kewley},
  {Graham}, \& {Fruchter}}]{Levesque10c}
{Levesque}, E.~M., {Kewley}, L.~J., {Graham}, J.~F., \& {Fruchter}, A.~S.
  2010{\natexlab{c}}, \apjl, 712, L26

\bibitem[{{Levesque} {et~al.}(2010{\natexlab{d}}){Levesque}, {Kewley}, \&
  {Larson}}]{Levesque10a}
{Levesque}, E.~M., {Kewley}, L.~J., \& {Larson}, K.~L. 2010{\natexlab{d}}, \aj,
  139, 712

\bibitem[{{Levesque} {et~al.}(2010{\natexlab{e}}){Levesque}, {Soderberg},
  {Kewley}, \& {Berger}}]{Levesque10e}
{Levesque}, E.~M., {Soderberg}, A.~M., {Kewley}, L.~J., \& {Berger}, E.
  2010{\natexlab{e}}, ArXiv e-prints

\bibitem[{{Maiolino} {et~al.}(2008){Maiolino}, {Nagao}, {Grazian}, {Cocchia},
  {Marconi}, {Mannucci}, {Cimatti}, {Pipino}, {Ballero}, {Calura}, {Chiappini},
  {Fontana}, {Granato}, {Matteucci}, {Pastorini}, {Pentericci}, {Risaliti},
  {Salvati}, \& {Silva}}]{Maiolino08}
{Maiolino}, R., {Nagao}, T., {Grazian}, A., {et~al.} 2008, \aap, 488, 463

\bibitem[{{Malesani} {et~al.}(2009){Malesani}, {Hjorth}, {Fynbo},
  {Milvang-Jensen}, {Jakobsson}, \& {Jaunsen}}]{Malesani09}
{Malesani}, D., {Hjorth}, J., {Fynbo}, J.~P.~U., {et~al.} 2009, in American
  Institute of Physics Conference Series, Vol. 1111, American Institute of
  Physics Conference Series, ed. {G.~Giobbi, A.~Tornambe, G.~Raimondo,
  M.~Limongi, L.~A.~Antonelli, N.~Menci, \& E.~Brocato}, 513--519

\bibitem[{{Mannucci} {et~al.}(2010){Mannucci}, {Cresci}, {Maiolino}, {Marconi},
  \& {Gnerucci}}]{Mannucci10}
{Mannucci}, F., {Cresci}, G., {Maiolino}, R., {Marconi}, A., \& {Gnerucci}, A.
  2010, \mnras, 408, 2115

\bibitem[{{Mannucci} {et~al.}(2009){Mannucci}, {Cresci}, {Maiolino}, {Marconi},
  {Pastorini}, {Pozzetti}, {Gnerucci}, {Risaliti}, {Schneider}, {Lehnert}, \&
  {Salvati}}]{Mannucci09b}
{Mannucci}, F., {Cresci}, G., {Maiolino}, R., {et~al.} 2009, \mnras, 398, 1915

\bibitem[{{Mao} \& {Mo}(1998)}]{Mao98}
{Mao}, S. \& {Mo}, H.~J. 1998, \aap, 339, L1

\bibitem[{{Mobasher} {et~al.}(2009){Mobasher}, {Dahlen}, {Hopkins}, {Scoville},
  {Capak}, {Rich}, {Sanders}, {Schinnerer}, {Ilbert}, {Salvato}, \&
  {Sheth}}]{Mobasher09}
{Mobasher}, B., {Dahlen}, T., {Hopkins}, A., {et~al.} 2009, \apj, 690, 1074

\bibitem[{{Modjaz} {et~al.}(2008){Modjaz}, {Kewley}, {Kirshner}, {Stanek},
  {Challis}, {Garnavich}, {Greene}, {Kelly}, \& {Prieto}}]{Modjaz08}
{Modjaz}, M., {Kewley}, L., {Kirshner}, R.~P., {et~al.} 2008, \aj, 135, 1136

\bibitem[{{Nagao} {et~al.}(2006){Nagao}, {Maiolino}, \& {Marconi}}]{Nagao06}
{Nagao}, T., {Maiolino}, R., \& {Marconi}, A. 2006, \aap, 459, 85

\bibitem[{{Osterbrock}(1989)}]{Osterbrock89}
{Osterbrock}, D.~E. 1989, {Astrophysics of gaseous nebulae and active galactic
  nuclei} ({Univ. Science books})

\bibitem[{{Perley} {et~al.}(2009){Perley}, {Cenko}, {Bloom}, {Chen}, {Butler},
  {Kocevski}, {Prochaska}, {Brodwin}, {Glazebrook}, {Kasliwal}, {Kulkarni},
  {Lopez}, {Ofek}, {Pettini}, {Soderberg}, \& {Starr}}]{Perley09b}
{Perley}, D.~A., {Cenko}, S.~B., {Bloom}, J.~S., {et~al.} 2009, \aj, 138, 1690

\bibitem[{{Pettini} \& {Pagel}(2004)}]{Pettini04}
{Pettini}, M. \& {Pagel}, B.~E.~J. 2004, \mnras, 348, L59

\bibitem[{{Porciani} \& {Madau}(2001)}]{Porciani01}
{Porciani}, C. \& {Madau}, P. 2001, \apj, 548, 522

\bibitem[{{Price} {et~al.}(2007){Price}, {Songaila}, {Cowie}, {Bell Burnell},
  {Berger}, {Cucchiara}, {Fox}, {Hook}, {Kulkarni}, {Penprase}, {Roth}, \&
  {Schmidt}}]{Price07}
{Price}, P.~A., {Songaila}, A., {Cowie}, L.~L., {et~al.} 2007, \apjl, 663, L57

\bibitem[{{Prochaska} {et~al.}(2004){Prochaska}, {Bloom}, {Chen}, {Hurley},
  {Melbourne}, {Dressler}, {Graham}, {Osip}, \& {Vacca}}]{Prochaska04}
{Prochaska}, J.~X., {Bloom}, J.~S., {Chen}, H., {et~al.} 2004, \apj, 611, 200

\bibitem[{{Salvaterra} {et~al.}(2009){Salvaterra}, {Della Valle}, {Campana},
  {Chincarini}, {Covino}, {D'Avanzo}, {Fern{\'a}ndez-Soto}, {Guidorzi},
  {Mannucci}, {Margutti}, {Th{\"o}ne}, {Antonelli}, {Barthelmy}, {de Pasquale},
  {D'Elia}, {Fiore}, {Fugazza}, {Hunt}, {Maiorano}, {Marinoni}, {Marshall},
  {Molinari}, {Nousek}, {Pian}, {Racusin}, {Stella}, {Amati}, {Andreuzzi},
  {Cusumano}, {Fenimore}, {Ferrero}, {Giommi}, {Guetta}, {Holland}, {Hurley},
  {Israel}, {Mao}, {Markwardt}, {Masetti}, {Pagani}, {Palazzi}, {Palmer},
  {Piranomonte}, {Tagliaferri}, \& {Testa}}]{Salvaterra09}
{Salvaterra}, R., {Della Valle}, M., {Campana}, S., {et~al.} 2009, \nat, 461,
  1258

\bibitem[{{Salvaterra} {et~al.}(2010){Salvaterra}, {Ferrara}, \&
  {Dayal}}]{Salvaterra10}
{Salvaterra}, R., {Ferrara}, A., \& {Dayal}, P. 2010, ArXiv e-prints 1003.3873

\bibitem[{{Savaglio} {et~al.}(2007){Savaglio}, {Budav{\'a}ri}, {Glazebrook},
  {Le Borgne}, {Le Floc'h}, {Chen}, {Greiner}, \& {Yoldas}}]{Savaglio07}
{Savaglio}, S., {Budav{\'a}ri}, T., {Glazebrook}, K., {et~al.} 2007, The
  Messenger, 128, 47

\bibitem[{{Savaglio} {et~al.}(2005){Savaglio}, {Glazebrook}, {Le Borgne},
  {Juneau}, {Abraham}, {Chen}, {Crampton}, {McCarthy}, {Carlberg}, {Marzke},
  {Roth}, {J{\o}rgensen}, \& {Murowinski}}]{Savaglio05}
{Savaglio}, S., {Glazebrook}, K., {Le Borgne}, D., {et~al.} 2005, \apj, 635,
  260

\bibitem[{{Savaglio} {et~al.}(2009){Savaglio}, {Glazebrook}, \&
  {LeBorgne}}]{Savaglio09}
{Savaglio}, S., {Glazebrook}, K., \& {LeBorgne}, D. 2009, \apj, 691, 182

\bibitem[{{Stanek} {et~al.}(2006){Stanek}, {Gnedin}, {Beacom}, {Gould},
  {Johnson}, {Kollmeier}, {Modjaz}, {Pinsonneault}, {Pogge}, \&
  {Weinberg}}]{Stanek06}
{Stanek}, K.~Z., {Gnedin}, O.~Y., {Beacom}, J.~F., {et~al.} 2006, \actaa, 56,
  333

\bibitem[{{Svensson} {et~al.}(2010){Svensson}, {Levan}, {Tanvir}, {Fruchter},
  \& {Strolger}}]{Svensson10}
{Svensson}, K.~M., {Levan}, A.~J., {Tanvir}, N.~R., {Fruchter}, A.~S., \&
  {Strolger}, L. 2010, \mnras, 405, 57

\bibitem[{{Tanvir} {et~al.}(2009){Tanvir}, {Fox}, {Levan}, {Berger},
  {Wiersema}, {Fynbo}, {Cucchiara}, {Kr{\"u}hler}, {Gehrels}, {Bloom},
  {Greiner}, {Evans}, {Rol}, {Olivares}, {Hjorth}, {Jakobsson}, {Farihi},
  {Willingale}, {Starling}, {Cenko}, {Perley}, {Maund}, {Duke}, {Wijers},
  {Adamson}, {Allan}, {Bremer}, {Burrows}, {Castro-Tirado}, {Cavanagh}, {de
  Ugarte Postigo}, {Dopita}, {Fatkhullin}, {Fruchter}, {Foley}, {Gorosabel},
  {Kennea}, {Kerr}, {Klose}, {Krimm}, {Komarova}, {Kulkarni}, {Moskvitin},
  {Mundell}, {Naylor}, {Page}, {Penprase}, {Perri}, {Podsiadlowski}, {Roth},
  {Rutledge}, {Sakamoto}, {Schady}, {Schmidt}, {Soderberg}, {Sollerman},
  {Stephens}, {Stratta}, {Ukwatta}, {Watson}, {Westra}, {Wold}, \&
  {Wolf}}]{Tanvir09}
{Tanvir}, N.~R., {Fox}, D.~B., {Levan}, A.~J., {et~al.} 2009, \nat, 461, 1254

\bibitem[{{Totani}(1997)}]{Totani97}
{Totani}, T. 1997, \apjl, 486, L71+

\bibitem[{{Totani} {et~al.}(2006){Totani}, {Kawai}, {Kosugi}, {Aoki}, {Yamada},
  {Iye}, {Ohta}, \& {Hattori}}]{Totani06}
{Totani}, T., {Kawai}, N., {Kosugi}, G., {et~al.} 2006, \pasj, 58, 485

\bibitem[{{Tremonti} {et~al.}(2004){Tremonti}, {Heckman}, {Kauffmann},
  {Brinchmann}, {Charlot}, {White}, {Seibert}, {Peng}, {Schlegel}, {Uomoto},
  {Fukugita}, \& {Brinkmann}}]{Tremonti04}
{Tremonti}, C.~A., {Heckman}, T.~M., {Kauffmann}, G., {et~al.} 2004, \apj, 613,
  898

\bibitem[{{Watson} {et~al.}(2006){Watson}, {Fynbo}, {Ledoux}, {Vreeswijk},
  {Hjorth}, {Smette}, {Andersen}, {Aoki}, {Augusteijn}, {Beardmore}, {Bersier},
  {Castro Cer{\'o}n}, {D'Avanzo}, {Diaz-Fraile}, {Gorosabel}, {Hirst},
  {Jakobsson}, {Jensen}, {Kawai}, {Kosugi}, {Laursen}, {Levan}, {Masegosa},
  {N{\"a}r{\"a}nen}, {Page}, {Pedersen}, {Pozanenko}, {Reeves}, {Rumyantsev},
  {Shahbaz}, {Sharapov}, {Sollerman}, {Starling}, {Tanvir}, {Torstensson}, \&
  {Wiersema}}]{Watson06}
{Watson}, D., {Fynbo}, J.~P.~U., {Ledoux}, C., {et~al.} 2006, \apj, 652, 1011

\bibitem[{{Wijers} {et~al.}(1998){Wijers}, {Bloom}, {Bagla}, \&
  {Natarajan}}]{Wijers98}
{Wijers}, R.~A.~M.~J., {Bloom}, J.~S., {Bagla}, J.~S., \& {Natarajan}, P. 1998,
  \mnras, 294, L13

\bibitem[{{Wolf} \& {Podsiadlowski}(2007)}]{Wolf07}
{Wolf}, C. \& {Podsiadlowski}, P. 2007, \mnras, 375, 1049

\bibitem[{{Woosley} \& {Heger}(2006)}]{Woosley06}
{Woosley}, S.~E. \& {Heger}, A. 2006, \apj, 637, 914

\bibitem[{{Yoon} {et~al.}(2008){Yoon}, {Langer}, {Cantiello}, {Woosley}, \&
  {Glatzmaier}}]{yoon08}
{Yoon}, S., {Langer}, N., {Cantiello}, M., {Woosley}, S.~E., \& {Glatzmaier},
  G.~A. 2008, in IAU Symposium, Vol. 250, IAU Symposium, ed. {F.~Bresolin,
  P.~A.~Crowther, \& J.~Puls}, 231--236

\bibitem[{{Yoon} {et~al.}(2006){Yoon}, {Langer}, \& {Norman}}]{Yoon06}
{Yoon}, S., {Langer}, N., \& {Norman}, C. 2006, \aap, 460, 199

\bibitem[{{Zhang} \& {M{\'e}sz{\'a}ros}(2004)}]{Zhang04}
{Zhang}, B. \& {M{\'e}sz{\'a}ros}, P. 2004, International Journal of Modern
  Physics A, 19, 2385

\bibitem[{{Zhao} {et~al.}(2010){Zhao}, {Gao}, \& {Gu}}]{Zhao10}
{Zhao}, Y., {Gao}, Y., \& {Gu}, Q. 2010, \apj, 710, 663

\end{thebibliography}

\end{document}